# Selective High-Harmonic Generation by Charge Transfer Plasmon-Resonant Nanodimers

*Burak Gerislioglu\*[‡], Arash Ahmadivand[‡], and Nezih Pala*

*Department of Electrical and Computer Engineering, Florida International University, 10555 W Flagler St., Miami, FL 33174, USA*

\*Corresponding Author: bgeri002@fiu.edu

**Abstract-** Here, we study the possibility of selective high-harmonic generation (HHG) based on the concept of charge transfer plasmons (CTP) in bridged nanoparticle assemblies. By choosing plasmonic dimer nanoantenna, as a fundamental member of nanoclusters family, and bridging the gap space between the proximal nanoparticles with an optothermally controllable substance, we selectively produced two different harmonics at different states of the functional bridge. To this end, the conductive connection between the nanoparticles is mediated with $Ge_2Sb_2Te_5$ (GST), which inherently has opposite optical and electrical properties depending on its phase condition. This helped us to manipulate the transition of charges across the bridge and therefore, control the excitation of CTP resonances and provide a switching feature between distinguished plasmonic resonant modes. Such a control allows us to produce highly intense second and forth harmonics based on the conductivity of the interparticle junction.

[‡]: equal contribution.

**Introduction**

Nonlinear plasmonics is one of the fundamental counterparts of optical and photonic sciences with a wide range of fascinating applications in development of integrated nanolasers, advanced light sources, and next-generation subwavelength devices [1-3]. An intense localization of optically driven hot electrons in plasmonic structures in the associated resonant frequency can substantially enhance the nonlinear optical effects, leading to have an active control over the light in subwavelength regime [4]. As a conventional nonlinear phenomenon, high harmonic generation (HHG) has been observed, experienced, and reported in gas atoms such as Argon (Ar) and Xenon (Xe), due to multiphoton ionization [5]. To date, generation of higher-order harmonics have been achieved using various systems such as nanophotonic and plasmonic structures [4]. Here the major idea is based on the conversion of fundamental incident beam frequency into the intense higher-order harmonics by nanoscale bulk solids with large field-induced of susceptibilities ($\chi^{(2)}$ or $\chi^{(4)}$) [1-3,4,6].

One of the successful and promising approaches for HHG is the utilization of resonant nanostructures including both plasmonic and high-index all-dielectric systems. Relatively, for instance, Fano-resonant nanostructures [7-10] are the most popular selection to produce nonlinear properties and generate high-order moments. Taking the advantage of strong hotspots and near-field coupling in Fano-resonant systems, highly intense nonlinear responses have been achieved [11,12]. Although there have been significant achievements in nonlinear plasmonics, these systems suffer from poor selectivity. In other words, generation of both odd and even harmonics using a single system would help to develop miniature lasers and functional nonlinear metadevices. To address this limitation, graphene monolayer with the ability to tune the plasmonic response based on control over the carrier concentration have been proposed for extreme nonlinear photonics and HHG using the Kerr nonlinearity [13,14]. However, graphene-based nanostructures suffer from complex and expensive fabrication techniques, and lack of mechanical flexibility [15,16].

Newly, a new spectral feature has been introduced, known as charge transfer plasmons (CTP), excitable in plasmonic nanostructures with subnanometer atomic-scale openings [17,18], and also in nanoparticles with conductive junctions [19-21]. In the earlier method, the quantum tunneling plays a key role for the excitation of CTPs, has been described by quantum corrected model. On the other hand, for the latter case, the direct transition of charges across the nanobridge between conductive nanoparticles acts as a fundamental component for the excitation of CTPs. So far, various methods have been introduced to increase the inherent and limited tunability of CTPs such as using phase-change materials (PCMs) (i.e. $Ge_2Sb_2Te_5$ and $VO_2$) [22-25] and graphene [26,27] monolayer to control the transfer of charges and tune the CTP feature. Tunable CTPs have been recently employed for developing ultrafast telecommunication devices such as switches [27] and metamodulators [7].

In this study, we propose the first study over the use of optothermally tunable CTP resonant-nanodimers for the selective generation of high-order even harmonics ($2\omega$ and $4\omega$). To this end, we used a simple nanodimer composed of two nanodisks connected with a metallodielectric junction consisting of metallic and PCM (here $Ge_2Sb_2Te_5$ or simply GST) blocks to each other. Particularly, as a well-known member of PCM family, GST possesses two opposite behaviors below (amorphous) and above (crystalline) 477 °C [28]. Such an exquisite feature allows for controlling the transition of charges across the bridge, therefore, tune the generated nonlinear response and select the targeted higher-order harmonic. We have previously employed this exquisite feature for plasmonic switching application at the telecommunication band [20]. We showed that applying either optical or external heat stimuli facilitates an interplay between the opposite phases of the GST bridge and gives rise to fast switching between all-dielectric and conductive regimes. This results with conversion between the dipolar and CTP moments at different wavelengths. For the nonlinear application and when the GST-mediated nanobridge is in dielectric regime, the capacitive coupling is dominant and the dipolar extreme appears at $\lambda \sim 1800$ nm. Conversely, when the GST section turns into the crystalline state, the bridge acts as a fully-metallic junction, and gives rise to formation of a CTP peak around $\lambda \sim 2100$ nm. This feature allows for developing selective HHG depending on the GST state.

**Results and Discussions**

Figure 1a demonstrates a schematic of the proposed metallodielectric dimer (not to scale). The geometrical parameters are specified in Fig. 1b. The following dimensions are utilized for the structure: $W/L/D$= 50, 30, and 200 nm, respectively. Besides, the total length of the compositional bridge is fixed to 200 nm. It is assumed that the dimer is deposited on a glass ($SiO_2$) substrate with Palik permittivity constants [29]. Here, the nanodisks are gold (Au) with Johnson and Christy constants [30]. The complex permittivity values for GST at both phases were taken from the experimental results reported by Shportko *et al*. [31]. In our electromagnetic analyses, we presumed that GST and Au parts of the nanobridge are perfectly matched, and the thickness is homogenous and fixed to $t$= 45 nm.

Here, we focus on extracting the spectral response of the metallodielectric dimer under *x*-polarized beam illumination for three different states of the bridge: dielectric (a-GST), conductive (c-GST), and fully metallic regimes. The normalized extinction cross-sections for the dimer are exhibited in Fig. 2a. For the full Au junction, the structure acts as a classical CTP-resonant system [32], and a pronounced CTP mode formed around $\lambda \sim 2400$ nm due to free transition of the charges across the conductive wire. The excited CTP peak for the fully metallic regime is not tunable and morphological variations must be applied to tune the spectral feature. By introducing a small section of GST at the center of Au nanobridge, an efficiently tuning of the plasmonic response of the nanoassembly would be possible. To this end, initially, we assumed that GST is in the amorphous state and acting similar to a dielectric spacer. This results in formation of a classical dipole at $\lambda$~1800 nm due to strong capacitive coupling between the proximal nanostructures. Switching the state of GST to crystalline regime, we facilitate the transition of charges instead of coupling. The low resistivity of the c-GST (~$10^{-3}$ Ω.cm) leads to the formation of a CTP peak around $\lambda$~2100 nm. It should be noted that a shoulder appeared in all analyzed regimes around $\lambda$~650 nm correlating with the dipolar moment. The local electric field (E-field) density maps across the nanoassembly are plotted in Figs. 2b and 2c, using both FDTD and FEM tools. Obviously, by moving from a-GST (Figs. 2b,c(i)) to c-GST (Figs. 2b,c(ii)) regime, the capacitive coupling switches to the direct charge transfer and therefore, dipolar moment converts to the CTP mode. Relatively, for the dimer with fully metallic junction, the induced charges easily travel along the junction and concentrate in the outermost nanodisks (Figs. 2b,c(iii)).

In continue, we study the nonlinear response of the proposed functional nanodimer for selective generation of higher-order harmonics. To this end, we use the obtained fundamental resonance frequencies for both dipolar in a-GST and CTP in c-GST regimes of the bridge, and also we assume that the dominant contribution in generation of higher harmonics is from the metallodielectric nanodimer [11]. By applying a time-dependent signal with the pulse length of 2 ns (see Methods), we extracted the corresponding HHG intensity intensities in Figs. 3 and 4 for the bridge with a-GST and c-GST blocks, respectively. As shown in Fig. 3a, in the a-GST limit and for the beam with the fundamental wavelength of $\lambda$~1800 nm ($\omega$) for the dipolar moment, we observed formation of strong second harmonic at $\lambda$~980 nm ($2\omega$). It should be underlined that there is an insignificant deviation around 80 nm, since we neglected the effect of second harmonic susceptibility dispersion. The geometrical variations in the structural parameters of the dimer reveal strong dependency of the generated second harmonic on the corresponding dimensions. Figures 3b and 3c illustrate the variations in the length and width of the bridge, respectively, and the intensity profile for the variations in the thickness of the nanoplasmonic system is depicted in Fig. 3d. These results validate the highest SHG intensity of 4500 for the following geometries: with the size of *W*/*L*/*t*=50/30/45 nm. Technically, when the length of GST is increased from 30 nm to 50 nm and 75 nm, the wavelength of this harmonic blue-shifts to $\lambda$~730 nm, then red-shifts to $\lambda$~1200 nm, respectively. Further increases in length (to 100 nm) blue-shift the second harmonic to the earlier position, while the intensity reduces. For the bridge width variations in the range of 40 nm, 60 nm and 70 nm, we observed a dramatic reduction in the intensity of the generated harmonic as well as significant deviations in the position of second harmonic. As a final parameter, when the corresponding thickness (*t*) is varying from 45 nm to 60 nm, 90 nm, and 120 nm, the generated harmonic significantly shifts and the amplitude of the intensity reduces in comparison to the optimal dimensions.

On the other hand, for the metallodielectric dimer with c-GST nanobridge, an intense forth-harmonic ($4\omega$) is generated at $\lambda$~530 nm in comparison to its primary wavelength, $\lambda$~2100 nm ($\omega$), as plotted in Fig. 4a. Similar to a-GST high-harmonic generation analysis, here also we monitored a deviation between the simulated and expected fourth-harmonic around 5 nm. The effect of the geometrical variations on the FHG are presented in Figs. 4b-4d. Here, Increasing the length of GST section from 30 nm to 50 nm and 75 nm, the wavelength of generated harmonic remains constant, then red-shifts to $\lambda$~715 nm and $\lambda$~755 nm, respectively, while the intensity slightly decreases. In addition, increasing the width of the bridge in the range of 40 nm to 60 nm leads to dramatic decays in the intensity of the induced harmonic and blue-shifting to $\lambda$~250 nm and $\lambda$~280 nm, respectively. Eventually, while the thickness changes from 45 nm to 60 nm, 90 nm, and 120 nm, the position of the generated harmonic blue-shifts and the intensity decreased.

In Fig. 5, we estimated the quality-factor ($Q$-factor) [33-36] for both SHG and FHG as a function of geometrical variations for the GST block. The $Q$-factors for both phases are defined as ~6500 (c-GST) and ~450 (a-GST), respectively. Noticing in Fig. 5a, by increasing $W$ up to 70 nm and decreasing $L$ down to 50 nm, the $Q$-factor increases for the a-GST. The inset shows the $Q$-factor variations as a function of $t$. Obviously, fixing the $t$ value to 90 nm leads to a minor increase in the $Q$-factor. For the c-GST, by decreasing $L$ down to 30 nm and increasing $W$ up to 50 nm, the $Q$-factor increases substantially. Considering the inset in Fig. 5b, by reducing $t$ from 45 nm to 60 nm the $Q$-factor reduces, and further increases in $t$ slightly change the $Q$-factor. Ultimately, we quantified the conversion efficiency of the proposed structure. For a fixed beam power, the associated SHG and FHG conversion efficiencies define by: $\eta_{SHG} = P(2\omega)/P(\omega)$ and $\eta_{FHG} = P(4\omega)/P(\omega)$, and the corresponding values are 0.0095% and 0.017%, respectively.

## Conclusions

In conclusion, we have shown efficient HHG using optothermally controllable metallodielectric dimer. The proposed structure allowed us to induce two different types of plasmonics modes depending on the phase of GST, without requiring any geometrical changes. An active interplay between the amorphous and crystalline states of GST allowed for selective HHG with high conversion efficiency.

## Methods

The corresponding electromagnetic simulations of the proposed structure have been performed by using both finite-difference time domain (FDTD) (Lumerical 2018) and finite-element method (FEM) (COMSOL Multiphysics 5.2) methods. The incident beam radiation for crystallization was a broadband plane wave (400 nm-1600 nm), with the irradiation power of $P_0$= 3.2 μW, beam fluence of 60 Jm$^{-2}$, pulse duration of 500 fs, and repetition of 10 KHz. The boundaries of the workplace were surrounded with 64 layers of absorptive perfectly matched layers (PMLs). Besides, we utilized an additional light source with the duration of 0.9 ns and the irradiation power of 5.5 mW to provide the necessary thermal energy for the amorphization process. The spatial grid sizes with the size of 1 nm was applied, and the Courant stability was satisfied by setting the time step to *dt* ~0.1 fs [37]. Further, to obtain the generation of higher order harmonics, we utilized a time domain-based plane wave with the fundamental frequency based on the position of dipole or CTP mode, pulse length and offset of 2000 fs and 4000 fs, respectively.

## Acknowledgements

This work is supported by Army Research Laboratory (ARL) Multiscale Multidisciplinary Modeling of Electronic Materials (MSME) Collaborative Research Alliance (CRA) (Grant No. W911NF-12-2-0023, Program Manager: Dr. Meredith L. Reed). Arash Ahmadivand gratefully acknowledges the financial support provided through dissertational year fellowship (DYF) by the University Graduate School (UGS) at Florida International University.

## References

1  J. Y. Suh, C. H. Kim, W. Zhou, M. D. Huntington, D. T. Co, M. R. Wasielewski, and T. W. Odom, Plasmonic bowtie nanolaser arrays, *Nano Lett.* 12, 5769-5774 (2012).

2  B. Gerislioglu, A. Ahmadivand, and N. Pala, Hybridized plasmons in graphene nanorings for extreme nonlinear optics, *Opt. Mater.* 73, 729-735 (2017).

3  A. Ahmadivand, B. Gerislioglu, and N. Pala, Thermally controllable multiple high harmonics generation by phase-change materials-mediated Fano clusters, *arXiv Preprint arXiv: 1712.03802* (2017).

4  M. Kauranen, M., and A. V. Zayats, Nonlinear plasmonics, *Nature Photon.* 6, 737-748 (2012).

5  M. Ferray, A. L'Huillier, X. F. Li, L. A. Lompre, G. Mainfray, and C. Manus, Multiple-harmonic conversion of 1064 nm radiation in rare gases, *J. Phys. B: At. Mol. Opt. Phys.* 21, L31 (1988).

6  W. Cai, A. P. Vasudev, and M. L. Brongersma, Electrically controlled nonlinear generation of light with plasmonics, *Science* 333, 1720-1723 (2011).

7  A. Ahmadivand, B. Gerislioglu, and N. Pala, Active control over the interplay between the dark and hidden sides of plasmonics using metallodielectric Au-$Ge_2Sb_2Te_5$ unit cells, *J. Phys. Chem. C* 121, 19966-19974 (2017).

8  B. Gerislioglu, A. Ahmadivand, and N. Pala, Functional quadrumer clusters for switching between Fano and charge transfer plasmons, *IEEE Photon. Technol. Lett.* 29, 2226-2229 (2017).

9 B. Gerislioglu, A. Ahmadivand, and N. Pala, Single- and multimode beam propagation through an optothermally controllable Fano clusters-mediated waveguide, *IEEE J. Lightw. Technol*. 35, 4961-4966 (2017).

10 A. Ahmadivand, B. Gerislioglu, and N. Pala, Azimuthally and radially excited charge transfer plasmon and Fano lineshapes in conductive sublayer-mediated nanoassemblies, *J. Opt. Am. A* 34, 2052-2056 (2017).

11 K. Thyagarajan, J. Butet, and O. J. Martin, Augmenting second harmonic generation using Fano resonances in plasmonic systems, *Nano Lett.* 13, 1847-1851 (2013).

12 M.R. Shcherbakov, D.N. Neshev, B. Hopkins, A. S. Shorokhov, I. Staude, E. V. Melik-Gaykazyan, M. Decker, A. A. Ezhov, A. E. Miroshnichenko, I. Brener, and A. A. Fedyanin, Enhanced third-harmonic generation in silicon nanoparticles driven by magnetic response, *Nano Lett.* 14, 6488-6492 (2014).

13 S.-Y. Hong, J. I. Dadap, N. Petrone, P.-C. Yeh, J. Hone, and R. M. Osgood Jr., Optical third-harmonic generation in graphene, *Phys. Rev. X* 3, 021014 (2013).

14 J. D. Cox, A. Marini, and F. J. G. de Abajo, Plasmon-assisted high-harmonic generation in graphene, *Nature Commun*. 8, 14380 (2017).

15 A. Ahmadivand, R. Sinha, M. Karabiyik, P. K. Vabbina, B. Gerislioglu, S. Kaya, and N. Pala, Tunable THz wave absorption by graphene-assisted plasmonic metasurfaces based on metallic split ring resonators, *J. Nanopart. Res.* 19, 3 (2017)

16 A. Ahmadivand, M. Karabiyik, R. Sinha, B. Gerislioglu, and N. Pala, Tunable terahertz response of plasmonic vee-shaped assemblies with a graphene monolayer, *Proceedings of the IEEE Progress in Electromagnetic Research Symposium (PIERS)* 2387, Shanghai, China (2016), DOI:10.1109/PIERS.20167734979.

17 B. Gerislioglu, A. Ahmadivand, and N. Pala, Optothermally controlled charge transfer plasmons in Au-$Ge_2Sb_2Te_5$ core-shell assemblies, *arXiv preprint arXiv: 1712.01092* (2017).

18 S. F. Tan, L. Wu, J. K. Yang, P. Bai, M. Bosman, and C. A. Nijhuis, Quantum plasmon resonances controlled by molecular tunnel junctions, *Science* 343, 1496-1499 (2014).

19 A. Ahmadivand, R. Sinha, B. Gerislioglu, M. Karabiyik, N. Pala, and M. Shur, "Transition from capacitive coupling to direct charge transfer in asymmetric terahertz plasmonic assemblies," *Opt. Lett.* vol. 41, no. 22, pp. 5333-5336, Nov 2016.

20 A. Ahmadivand, B. Gerislioglu, R. Sinha, M. Karabiyik, and N. Pala, Optical switching using transition from dipolar to charge transfer plasmon modes in $Ge_2Sb_2Te_5$ bridged metallodielectric dimers, *Sci. Rep*. 7, 42807 (2017).

21 A. Ahmadivand, B. Gerislioglu, R. Sinha, P. K. Vabbina, M. Karabiyik, and N. Pala, Excitation of terahertz charge transfer plasmons in metallic fractal structures, *J. Infrared Milli. Terahz Waves* 38, 992- 1003 (2017).

22 G. Bakan, B. Gerislioglu, F. Dirisaglik, Z. Jurado, L. Sullivan, A. Dana, C. Lam, A. Gokirmak, and H. Silva, Extracting the temperature distribution on a phase-change memory cell during crystallization, *J. Appl. Phys.* 120, 164504 (2016).

23 V. Nooshnab, and A. Ahmadivand, Optothermally functional charge transfer plasmon modulator, *IEEE Photon. Technol. Lett.* 29, 1556-1559 (2017).

24 B. Gerislioglu, A. Ahmadivand, M. Karabiyik, R. Sinha, and N. Pala, $VO_2$-based reconfigurable antenna platform with addressable microheater matrix, *Adv. Electron. Mater.* 3, 1700170 (2017).

25 S. Raoux, H. Y. Cheng, M. A. Caldwell, and H.-S. P. Wong, Crystallization times of Ge-Te phase change materials as a function of composition, *Appl. Phys. Lett.* 95, 071910 (2009).

26 S. Thongrattanasiri, A. Manjavacas, P. Nordlander, and F. G. de Abajo, Quantum junction plasmons in graphene dimers, *Laser Photon. Rev.* 7, 297-302 (2013).

27 A. Ahmadivand, B. Gerislioglu, and N. Pala, Graphene optical switch based on charge transfer plasmons, *Phys. Status Solidi RRL* 11, 1700285 (2017).

28 J.-H. Kim, Correlation between microstructure and optical properties of $Ge_2Sb_2Te_5$ thin films, *J. Appl. Phys.* 86, 6770-6772 (1999).

29 E. D. Palik, *Handbook of Optical Constants of Solids*, San Diego, CA: Academic Press (1998).

30 P. B. Johnson, and R. W. Christy, Optical constants of the noble metals, *Phys. Rev. B* 6, 4370 (1972).

31 K. Shportko, S. Kremers, M. Woda, D. Lencer, J. Robertson, and M. Wuttig, Resonant bonding in crystalline phase change materials, *Nat. Mater.* 7, 653-658 (2008).

32 F. Wen, Y. Zhang, S. Gottheim, N. S. King, Y. Zhang, P. Nordlander, and N. J. Halas, Charge transfer plasmons: optical frequency conductances and tunable infrared resonances, *ACS Nano* 9, 6428-6435 (2015).

33 B. Gerislioglu, A. Ahmadivand, and N. Pala, Tunable plasmonic toroidal terahertz metamodulator, *arXiv preprint arXiv: 1712.04908* (2017).

34 A. Ahmadivand, B. Gerislioglu, and N. Pala, Large-modulation-depth polarization-sensitive plasmonic toroidal terahertz metamaterial, *IEEE Photon. Technol. Lett.* 29, 1860-1863 (2017).

35 A. Ahmadivand, B. Gerislioglu, P. Manickam, A. Kaushik, S. Bhansali, M. Nair, and N. Pala, Rapid detection of infectious envelope proteins by magnetoplasmonic toroidal metasensors, *ACS Sens.* 2, 1359-1368 (2017).

36 A. Ahmadivand, B. Gerislioglu, A. Tomitaka, P. Manickam, A. Kaushik, S. Bhansali, M. Nair, and N. Pala, Extreme sensitive metasensor for targeted biomarkers identification using colloidal nanoparticles-integrated plasmonic unit cells, *Biomed. Opt. Express* 9, 373-386 (2018).

37 F. Zheng, Z. Chen, and J. Zhang, A finite-difference time-domain method without the Courant stability conditions, *IEEE Microw. Guided Wave Lett.* 9, 441-443 (1999).

# Figures Captions

**Figure 1.** **(a)** and **(b)** Schematic and top-view pictures of the proposed nanodimer for HHG, respectively.

**Figure 2.** **(a)** Simulated extinction spectra of plasmonic dimer for the junction with full Au, a-GST and c-GST sections. **(b)** Surface charge density plots for the plasmonic dimer (FDTD): (i) dipolar mode (a-GST), (ii) CTP mode (c-GST), and CTP mode (Au). **(c)** Surface charge density maps for the dimer (FEM): (i) dipolar mode (a-GST), (ii) CTP mode (c-GST), and CTP mode (Au).

**Figure 3.** **(a)** The second-harmonic emission intensity response of the a-GST dimer. Relative emission intensity variations for **(b)** length ($L$), **(c)** width ($W$), **(d)** thickness ($t$).

**Figure 4.** **(a)** The fourth-harmonic emission intensity response of the c-GST dimer. Relative emission intensity variations for **(b)** length ($L$), **(c)** width ($W$), **(d)** thickness ($t$).

**Figure 5.** **(a)** SHG $Q$-factor as a function of length ($L$), width ($W$). The inset is the $Q$-factor of SHG as a function of thickness ($t$). **(b)** FHG $Q$-factor as a function of length ($L$), width ($W$). The inset is the $Q$-factor of FHG as a function of thickness ($t$).

Figure 1.

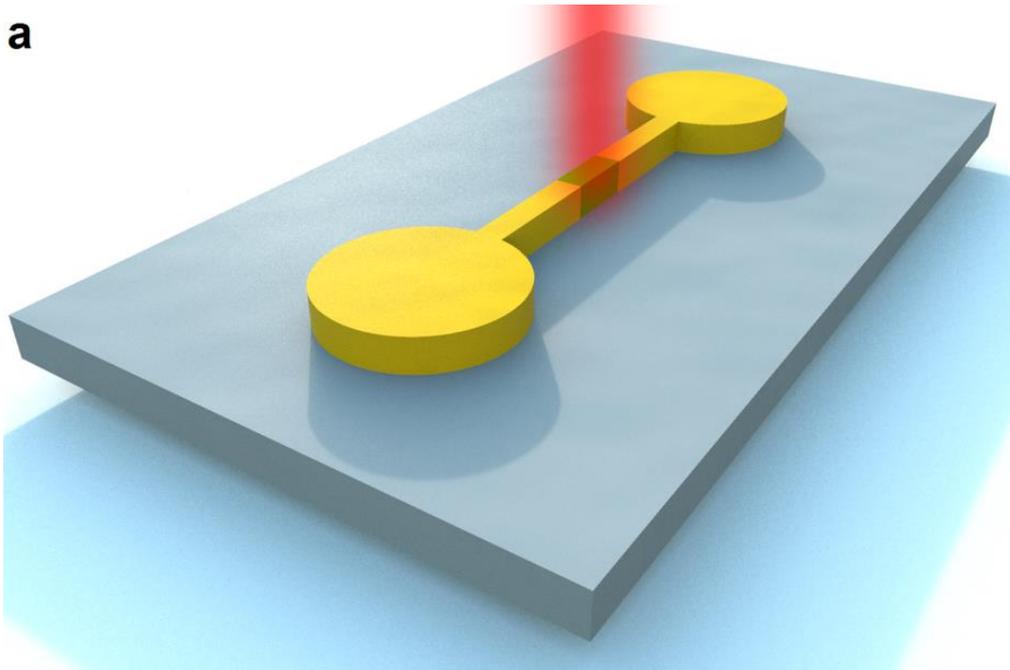

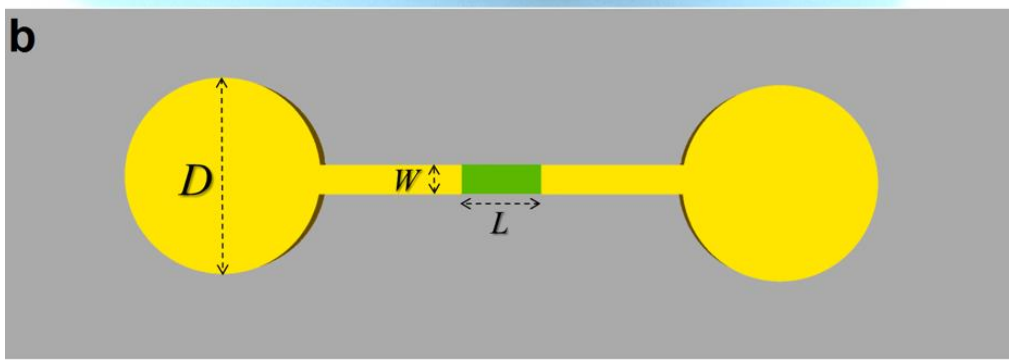

Figure 2.

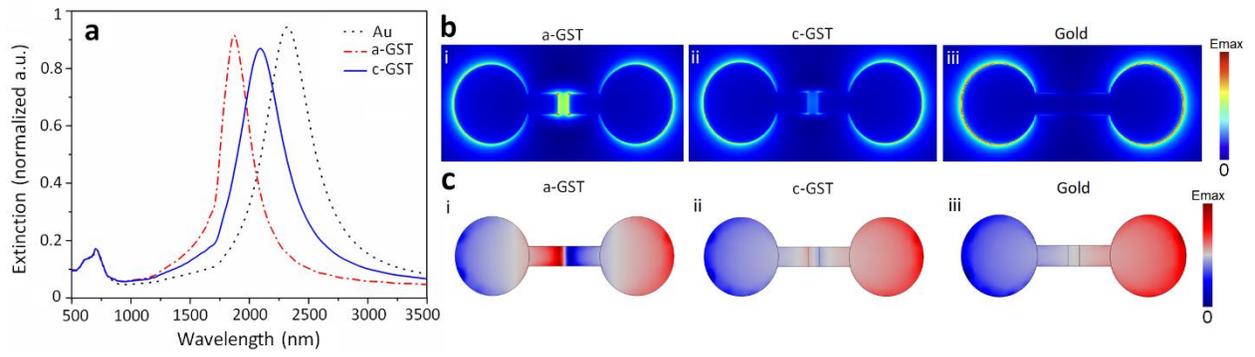

Figure 3.

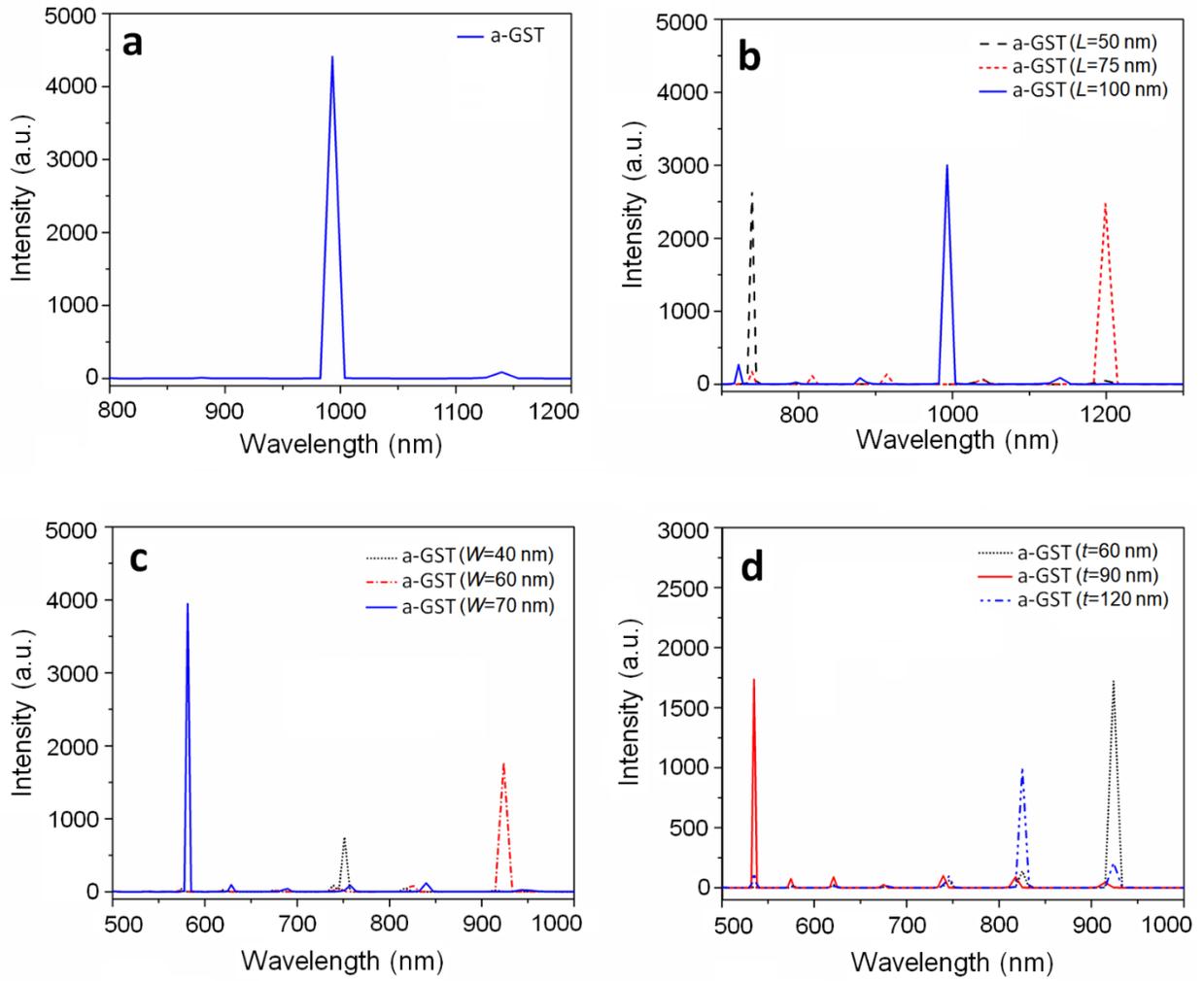

Figure 4.

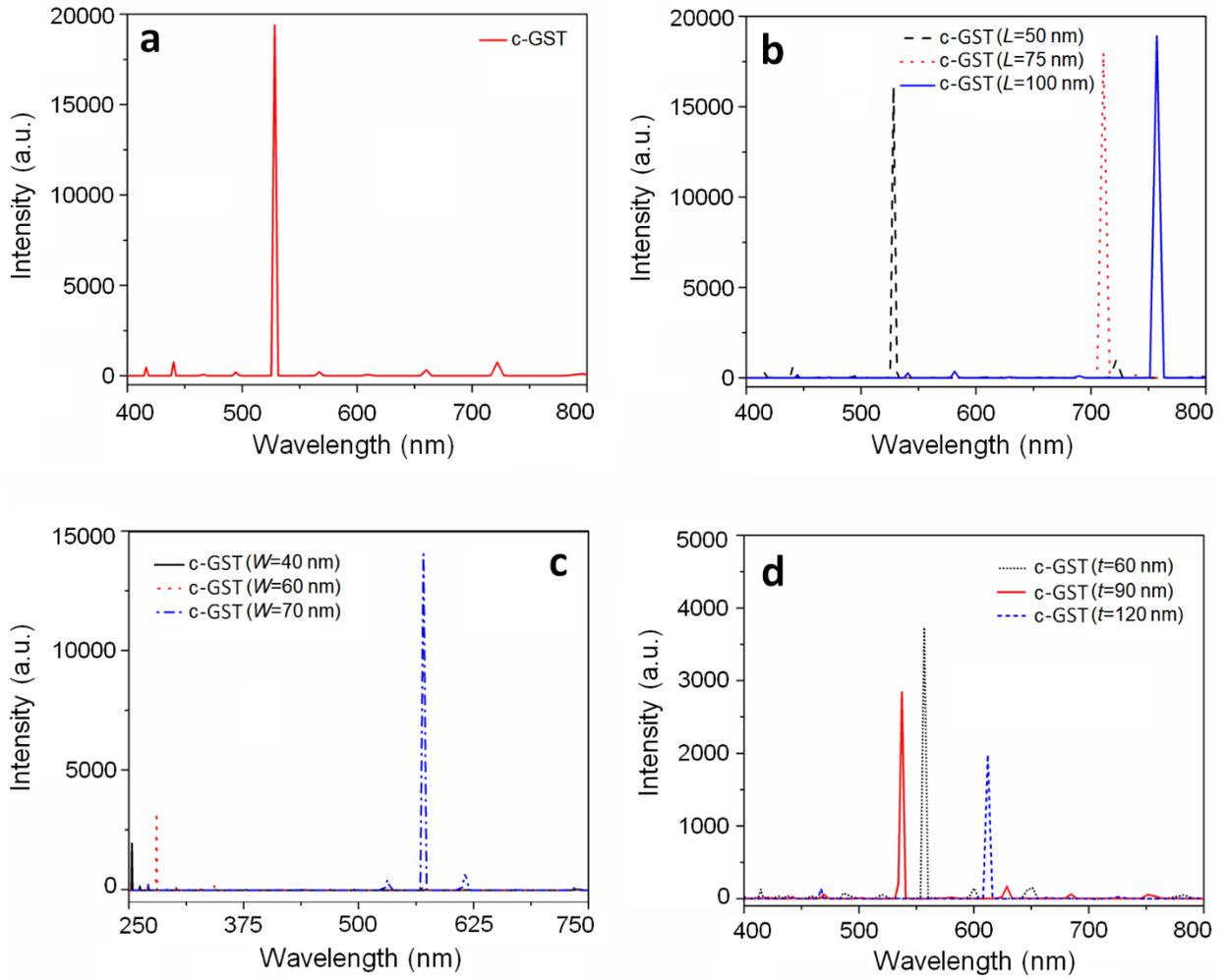

Figure 5.

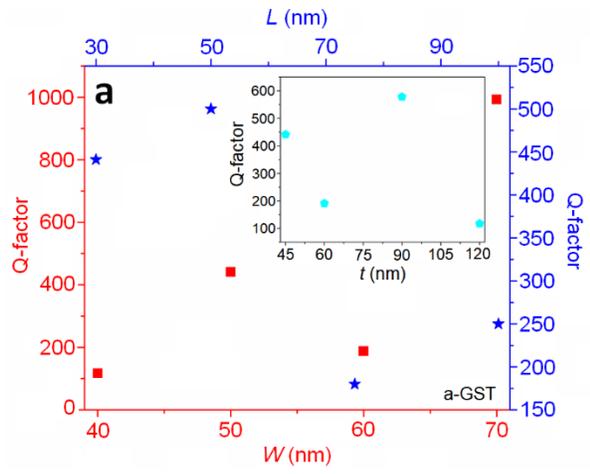 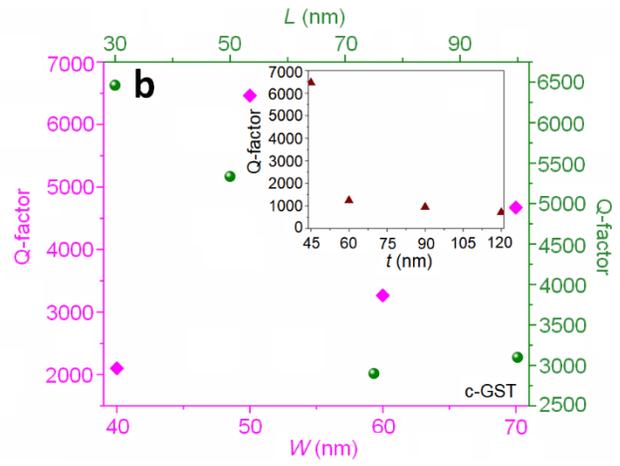